\PassOptionsToPackage{unicode}{hyperref}
\PassOptionsToPackage{hyphens}{url}
\PassOptionsToPackage{dvipsnames,svgnames,x11names}{xcolor}
\documentclass[
  letterpaper,
  DIV=11,
  numbers=noendperiod]{scrartcl}

\usepackage{amsmath,amssymb}
\usepackage{iftex}
\ifPDFTeX
  \usepackage[T1]{fontenc}
  \usepackage[utf8]{inputenc}
  \usepackage{textcomp} 
\else 
  \usepackage{unicode-math}
  \defaultfontfeatures{Scale=MatchLowercase}
  \defaultfontfeatures[\rmfamily]{Ligatures=TeX,Scale=1}
\fi
\usepackage{lmodern}
\ifPDFTeX\else  
\fi
\IfFileExists{upquote.sty}{\usepackage{upquote}}{}
\IfFileExists{microtype.sty}{
  \usepackage[]{microtype}
  \UseMicrotypeSet[protrusion]{basicmath} 
}{}
\makeatletter
\@ifundefined{KOMAClassName}{
  \IfFileExists{parskip.sty}{%
    \usepackage{parskip}
  }{
    \setlength{\parindent}{0pt}
    \setlength{\parskip}{6pt plus 2pt minus 1pt}}
}{
  \KOMAoptions{parskip=half}}
\makeatother
\usepackage{xcolor}
\usepackage{soul}
\ifx\paragraph\undefined\else
  \let\oldparagraph\paragraph
  \renewcommand{\paragraph}[1]{\oldparagraph{#1}\mbox{}}
\fi
\ifx\subparagraph\undefined\else
  \let\oldsubparagraph\subparagraph
  \renewcommand{\subparagraph}[1]{\oldsubparagraph{#1}\mbox{}}
\fi

\usepackage{longtable,booktabs,array}
\usepackage{calc} 
\usepackage{etoolbox}
\makeatletter
\patchcmd\longtable{\par}{\if@noskipsec\mbox{}\fi\par}{}{}
\makeatother
\IfFileExists{footnotehyper.sty}{\usepackage{footnotehyper}}{\usepackage{footnote}}
\makesavenoteenv{longtable}
\usepackage{graphicx}
\makeatletter
\def\maxwidth{\ifdim\Gin@nat@width>\linewidth\linewidth\else\Gin@nat@width\fi}
\def\maxheight{\ifdim\Gin@nat@height>\textheight\textheight\else\Gin@nat@height\fi}
\makeatother
\setkeys{Gin}{width=\maxwidth,height=\maxheight,keepaspectratio}
\makeatletter
\def\fps@figure{htbp}
\makeatother
\newlength{\cslhangindent}
\newlength{\csllabelwidth}
\newlength{\cslentryspacingunit} 
\newenvironment{CSLReferences}[2] 
 {
  \setlength{\parindent}{0pt}
  \ifodd #1
  \let\oldpar\par
  \def\par{\hangindent=\cslhangindent\oldpar}
  \fi
  \setlength{\parskip}{#2\cslentryspacingunit}
 }%
 {}
\usepackage{calc}

\KOMAoption{captions}{tableheading}
\usepackage{setspace}
\makeatletter
\@ifpackageloaded{caption}{}{\usepackage{caption}}
\AtBeginDocument{%
\ifdefined\contentsname
  \renewcommand*\contentsname{Table of contents}
\else
  \newcommand\contentsname{Table of contents}
\fi
\ifdefined\listfigurename
  \renewcommand*\listfigurename{List of Figures}
\else
  \newcommand\listfigurename{List of Figures}
\fi
\ifdefined\listtablename
  \renewcommand*\listtablename{List of Tables}
\else
  \newcommand\listtablename{List of Tables}
\fi
\ifdefined\figurename
  \renewcommand*\figurename{Figure}
\else
  \newcommand\figurename{Figure}
\fi
\ifdefined\tablename
  \renewcommand*\tablename{Table}
\else
  \newcommand\tablename{Table}
\fi
}
\@ifpackageloaded{float}{}{\usepackage{float}}
\floatstyle{ruled}
\@ifundefined{c@chapter}{\newfloat{codelisting}{h}{lop}}{\newfloat{codelisting}{h}{lop}[chapter]}
\floatname{codelisting}{Listing}

\makeatother
\makeatletter
\@ifpackageloaded{caption}{}{\usepackage{caption}}
\@ifpackageloaded{subcaption}{}{\usepackage{subcaption}}
\makeatother
\makeatletter
\@ifpackageloaded{tcolorbox}{}{\usepackage[skins,breakable]{tcolorbox}}
\makeatother
\makeatletter
\@ifundefined{shadecolor}{\definecolor{shadecolor}{rgb}{.97, .97, .97}}
\makeatother
\ifLuaTeX
  \usepackage{selnolig}  
\fi
\IfFileExists{bookmark.sty}{\usepackage{bookmark}}{\usepackage{hyperref}}
\IfFileExists{xurl.sty}{\usepackage{xurl}}{} 
\hypersetup{
  pdftitle={Creating Community in a Data Science Classroom},
  pdfauthor={David Kane},
  colorlinks=true,
  linkcolor={blue},
  filecolor={Maroon},
  citecolor={Blue},
  urlcolor={Blue},
  pdfcreator={LaTeX via pandoc}}

\title{Creating Community in a Data Science Classroom}
\author{David Kane}
\date{2023-09-01}

\begin{document}
\maketitle
\ifdefined\Shaded\renewenvironment{Shaded}{\begin{tcolorbox}[breakable, interior hidden, sharp corners, enhanced, borderline west={3pt}{0pt}{shadecolor}, boxrule=0pt, frame hidden]}{\end{tcolorbox}}\fi

\hypertarget{abstract}{%
\subsection*{Abstract}\label{abstract}}
\addcontentsline{toc}{subsection}{Abstract}

A community is a collection of people who know and care about each
other. The vast majority of college courses are not communities. This is
especially true of statistics and data science courses, both because our
classes are larger and because we are more likely to lecture. However,
it is possible to create a community in your classroom. This article
offers an idiosyncratic set of practices for creating community. I have
used these techniques successfully in first and second semester
statistics courses with enrollments ranging from 40 to 120. The key
steps are knowing names, cold calling, classroom seating, a shallow
learning curve, Study Halls, Recitations and rotating-one-on-one final
project presentations. *\textbf{Keywords}: education, data science.

\newpage

\hypertarget{names}{%
\subsection*{Names}\label{names}}
\addcontentsline{toc}{subsection}{Names}

Community starts with names. If two people don't know each other's
names, then it is hard to say that they really belong to the same
``community.'' The more that students know each other's names, the
tighter the classroom community will be.

Learn all your students' names. Is that easy? No! But all it takes is
time and concentration. Few of us can become more charismatic. All of us
can learn our students' names. Most schools have a system for sharing
student photos with their instructors. Make use of it. Of course,
students do change their appearance over time. Changes in hair style and
color are often tricky. As an instructor, I have two advantages. Since
students (see below) sit in the same area of the classroom each day, I
can use location as a cue. I can also study students during class while
they work on in-class assignments.

Use students' names. Greet them when they come into the lecture hall.
Will that freak them out? You bet! But it will also impress them, and
please them, although they will never admit that to you. All of us want
to be seen. Teach students each other's names. The typical lecture
classroom is a collection of strangers. Students come to class alone.
They sit alone, often in the same seat or at least region of the
classroom. If they have a friend in the class, they will often come into
the class with that friend, sit with that friend, and then leave with
that friend, never having interacted with another student in the class.
If they have a couple of friends in the class --- a common scenario for
members of a sports team --- they will travel and sit in a pack. They
don't mean to be unfriendly but other students will often perceive them
to be. Breaking down these barriers between students is the single most
important trick to creating a classroom community.

\hypertarget{seating}{%
\subsection*{Seating}\label{seating}}
\addcontentsline{toc}{subsection}{Seating}

Here are quotes from my syllabus, along with commentary:

\begin{quote}
Seating is organized, by campus geography, into several large ``Groups''
of 20 to 30 students: first years, Eliot House, Quadlings, et cetera.
Details depend on enrollment.
\end{quote}

Groups based on student housing are probably easiest, not least because
you want groups in which students are more likely to run into each other
outside of class. Groups based on class year are also sensible,
especially keeping all the first years together. The best approach will
depend on the details of your campus and student body. Each group will,
naturally, be diverse on dimensions other than the grouping criteria.

\begin{quote}
Students work in ``Pairs'' of two ``Partners.'' Sometimes, this will be
``side-by-side,'' each of you with a computer open, each writing code,
but talking with each other throughout. Other times, we will ``pair
program,'' meaning just one computer open and both of you collaborating
on a single project. You will work with a different partner every class.
\end{quote}

Students forced to work together for an entire class will have no choice
but to learn each other's names. Student culture changes over time, but
one aspect has been constant for decades. Almost all students wish they
knew more other students, and are happy to be introduced to them. At the
same time, few students will sit down next to a stranger in a classroom
and introduce themselves. Requiring students to work in pairs solves
this collective action problem. They want to meet each other. I require
students to work with different Partners each class because, without
that requirement, they won't.

\begin{quote}
If you are the stronger student in a Pair, do not simply charge ahead.
Instead, make sure that your Partner keeps up with you. Help each other!
If you aren't talking with each other often, then you are doing it
wrong. There is no better way to learn than to teach. The stronger
student should type less and talk more.
\end{quote}

There is nothing more exciting that a lecture hall with 50 conversations
going on simultaneously.

\begin{quote}
Besides your Partner, the students sitting immediately beside, behind
and in front of you are members of your Circle that day. Introduce
yourself to them when you/they arrive.
\end{quote}

Students won't want to do this. It will feel strange. Yet awkwardness in
the pursuit of community is no vice.

Sadly, this will only happen if you enforce it. Fortunately, enforcement
is easy. I begin each class with a random cold call and ask the lucky
student to introduce me to both her Partner and to the students in her
Circle. (If the first student fails at this, I will give the entire
class 30 seconds to do some quick introductions around their Circle
before I ask another student.) By the second week, students are doing
this on their own, as soon as they enter the classroom. This creates a
very different atmosphere. Recall the slogan for the TV show
\emph{Cheers}: ``Where everybody knows your name.'' A well-functioning
classroom community begins with everyone knowing your name, you knowing
theirs, and them knowing each other's.

\begin{quote}
Record the name of your Partner in the Google sheet for the day and the
names of your Circle in a different Google sheet. Each person does this,
even though doing so leads to duplication. (Don't stress about
spelling.)
\end{quote}

Requiring that names are recorded makes things easier for shy students.
They have no choice but to record the other students' names. It is a
course requirement. Each Google sheet is pre-formatted with all the
students in the class. Students will list Partner/Circle names next to
their own. This makes it easy for me to confirm that students are all
present, without wasting class time on calling attendance. If I notice a
student is missing --- which is easy to do with a glance at the sheet
--- I call on a member of their Group and inquire about the missing
student's well-being. When that student professes ignorance, as they
always will, I ask them to text/e-mail the missing student to ``make
sure she is OK.'' And I really do care about the health/safety of my
students! But I am also well-aware that the causal effect of this
practice is to maximize lecture attendance.

\hypertarget{cold-calling}{%
\subsection*{Cold Calling}\label{cold-calling}}
\addcontentsline{toc}{subsection}{Cold Calling}

Nothing keeps students engaged more than cold calling. These cold calls
are low pressure. Wrong answers do not matter. They are not counted in a
student's grade. To illustrate this, I often ask questions --- like
``What is my favorite soccer team?'' --- which students can't possibly
answer. Whenever a student is stumped, or even appears stumped, I
quickly answer the question myself or move on to the next student. But,
in general, questions are so simple, tied so directly to what the
students have just been working on, that students can easily answer. Yet
just the fact that they might be cold called ensures that they are all
paying attention, all the time.

I use an interactive R function to randomly select the student to call
on. Students can see my RStudio session, projected onto the big screen
at the front of the room, and watch me run the function. They know that
my cold calling is random, that there is no favoritism. (They are often
surprised at how non-random a truly random algorithm will appear, with
the same student being called on twice or even three times in a single
class session, even in a large class.) Cold calling becomes a bit of a
game, one in which students are both observers and participants.

\hypertarget{working-classroom}{%
\subsection*{Working classroom}\label{working-classroom}}
\addcontentsline{toc}{subsection}{Working classroom}

\begin{quote}
You learn how to play soccer with the ball at your feet. You learn how
to program with your hands on the keyboard.
\end{quote}

The term ``flipped classroom'' (Lemov (2015)) has two implications: the
first about what happens in the classroom and the second about what
happens outside the classroom. I prefer the term ``working classroom,''
because it references what goes in the classroom --- working not
lecturing --- without making claims about what should occur outside the
classroom.

In particular, lectures, whether given in class or required outside of
class, are among the worst methods for transferring information. First,
lectures (either in person or video) are too slow for almost 50\% of the
class. They have covered this topic in another class. They understood
this concept from the reading. You are wasting their time by explaining
X again. Second, lectures are too fast for almost 50\% of the class, for
the same reasons. By definition, a lecture can only be correctly paced
for, at most, a handful of students.

A ``working classroom'' is about what occurs in the class itself. For
me, this work is either programming or talking/writing about statistics.
Because students work in pairs, they are always working. They are always
either typing or directing what their partner should type. They are
always either talking, or listening to their partner. It is impossible
to not be engaged in a working classroom.

A working classroom creates a pit of success. Students can't help but to
learn something, even if it is only to practice a skill. (Soccer players
practice passing every day. Data scientists should practice using the
computer to work with data every day. You can always get better, even at
something you already ``know'' how to do.)

\hypertarget{no-student-left-behind}{%
\subsection*{No student left behind}\label{no-student-left-behind}}
\addcontentsline{toc}{subsection}{No student left behind}

The weakest students are most at risk for estrangement from the
classroom community. In every class, there will be strong students and
weak students. The vast majority of my focus as an instructor is devoted
to the weakest 20\% of the students, especially the bottom 5\%. Those
are the students who most need my help and are most likely to benefit
from it. Those are the students I want the teaching staff to take care
of. We should not ignore our stronger students, of course. But, at the
end of the semester, I judge myself most on the causal effect I have had
on the students who struggled most at the beginning.

First, I create the shallowest possible learning curve. Each step is as
small as possible. Once I have taught you A, I want B to be easy. Once
you understand B, C should be simple. And so on. All steps are baby
steps. By the end of the semester, we will have covered as much material
as a traditional class. Students learn as much, if not more. But,
instead of just problem sets every few weeks and a high pressure exam or
two, they have 30 or more assignments, each building on the previous
ones, all required.

Second, work should be spread out as much as possible. Learning
statistics is like learning a new language: you should practice every
day. In a typical class, students have tutorials due on Monday, several
hours of simple questions which are easy to do as long as you open the
textbook. It is hard to make students do the reading. It is easy to ask
them 100 questions which make it almost impossible for them not to do
the reading. Class on Tuesday (and Thursday) features 75 minutes of
intense collaborative data science work. Problem sets are due on
Wednesday. Students are encouraged to work together, to ask and answer
questions of each other and of course staff. Final project milestones
are due on Fridays. If you want students to work about 10 hours per week
outside of class, then they will learn the most if they spend 1 or 2
hours per day. Spreading out their work like this is not their natural
inclination. They need our help.

Third, students need to come to class. From my syllabus:

\begin{quote}
Missing Class: You expect me to be present for lecture. I expect the
same of you. There is nothing more embarrassing, for both us, than for
me to call your name and have you not be there to answer. But, at the
same time, conflicts arise. It is never a problem to miss class if, for
example, you are out of town or have a health issue. Just email me and
your assigned TF explaining the situation. Please do so on the day you
will be missing class. We don't need advanced warning.
\end{quote}

Note how cold calling provides a justification for enforcing lecture
attendance. Since the algorithm is random, nothing prevents it from
producing the name of a student who is not present. Of course, this
would not really be a problem in class, but it does provide an excuse
for insisting that students attend all classes, or inform us ahead of
time that they won't be present. The more often that students attend
class, the more that they will learn, the more that they will feel a
part of the classroom community.

\hypertarget{teaching-staff}{%
\subsection*{Teaching Staff}\label{teaching-staff}}
\addcontentsline{toc}{subsection}{Teaching Staff}

The larger the course, the more important the efforts of the teaching
staff toward nurturing a community. Colleges vary dramatically in the
types of teaching support they provide to data science courses. The raw
number of teaching staff, while almost always a function of the number
of students in the course, varies. Teaching staff can be anyone from
junior undergraduates to senior graduate students, even post-docs. The
titles and (permitted) duties of teaching staff often depends on their
undergraduate/graduate status. The number of hours is a function of both
the policies of the institution and the availability of the teaching
staff themselves. I will ignore that variation and address common
issues, referring to all teaching staff as teaching fellows (TFs).
Advice:

\begin{itemize}
\item
  Think in terms of hours rather than positions or roles. The total
  number of hours per week is the key resource, whether that is one TF
  who works 20 hours or four TFs who each work 5 hours. One key
  advantage of hours is that it is an institution-approved metric of
  workload. You may think that a specific TF is, for example,
  responsible for grading problem sets, but your institution does not
  use ``grade the problem sets'' as a metric. Another advantage of hours
  is that it helps to alleviate the principal-agent problem between you
  and your TFs. It is tough to ensure that TFs devote the correct amount
  of effort to their responsibilities. Specifying hours rather than
  tasks makes conflicts easier to manage.
\item
  Minimize time spent on grading. You and your TFs should automate it as
  much as possible. Services like Gradescope (Singh et al. (2017)) and
  PrairieLearn (West et al. (2015)) are helpful. Don't bother providing
  much written feedback, both because doing so is time consuming and
  because students often ignore it. Have a TF or two who specialize in
  grading. In that way, almost every hour that other TFs are paid for
  will involve time spent with students.
\item
  Make use of the beginning and the end of the semester. Many schools
  pay TFs for the entire semester, even for weeks before and after
  classes are actually meeting. Those are hours you can use even if
  other classes don't have student/TF meetings at those times.
\item
  Maximize the amount of time which TFs spend with students, either in
  small groups or one-on-one. Instead of (often optional) sections in
  which TFs lecture to students, arrange Recitations, small 30 or 60
  minute meetings between TFs and 1 to 4 students. I call these
  ``Recitations'' to highlight that they are different from the
  ``sections'' which students are used to. Use other terminology if you
  prefer.
\end{itemize}

\hypertarget{recitations}{%
\subsubsection*{Recitations}\label{recitations}}
\addcontentsline{toc}{subsubsection}{Recitations}

TFs should not attend your lectures. Although there are (maybe!)
benefits to having them in lecture, the opportunity cost is huge.
Instead of lurking in the back of the lecture hall, reading \st{Twitter}
\(\mathbb{X}\), for 2-3 hours per week, they could be meeting with small
groups of students.

TFs should not have (traditional) office hours. Most office hours are
unused by students. (Note that TFs have the incentive to schedule their
office hours at times and locations that students are less likely to
attend.)

Recitations are different from traditional sections for two reasons.
First, they involve much smaller groups. Instead of a single 60 minute
section with 20 students --- who might or might not attend, who might or
might not participate --- listening to a TF lecture, that same TF would
meet with students in groups of 4, for 60 minutes each. I am not
recommending that your TFs work more hours than they do now. They/you
are saving the 2-3 hours which they would have spent in lecture and the
1-2 hours they would have spent preparing their own lectures each week.
They spend those 3-5 extra hours with students in Recitations.

A community consists of people who care about one another. We want our
teaching staff to be invested in the success of their students. We want
our students to care about the opinions of the teaching staff, beyond
the brute cudgel provided by grading. The best way to create a
meaningful relationship between TFs and students is via hours spent
together, sitting around a table, talking about data science and,
ideally, working toward a common goal.

From the point of view of building community, the topic of the
Recitations is almost irrelevant. My recommendation is to focus those
meetings on final projects. A good structure is to have milestones for
your final projects due on Friday each week. The Recitation for that
week will focus on the TF helping students to complete the milestone.
There is nothing wrong with spending time answering questions or
discussing topics from class, but the main focus is the final project.
We want TFs and students to think of the projects as something they work
on together. We want TFs to be proud of their students when they present
their final projects. We want students to want to make their TFs proud.
Recitations make them care more about each other than they otherwise
would.

\hypertarget{study-halls}{%
\subsubsection*{Study Halls}\label{study-halls}}
\addcontentsline{toc}{subsubsection}{Study Halls}

The best replacement for office hours are Study Halls, 3 hour blocks of
time, located in a large space like a dining hall, hosted by a single
TF. From my instructions to students and teaching staff:

\begin{quote}
At every Study Hall, the TF will ensure that everyone knows everyone
else's name. These classes are communities and community begins with
names. The process starts with the first student arriving and sitting at
the table. They and the TF chat. (It is always nice for the student to
take the initiative and introduce themselves to the TF. Remembering all
your names is hard!) A second person arrives and sits at the same table,
followed by introductions. Persons 3 and 4 arrive. More introductions.
Help your TF by introducing yourself, even if you are 90\% sure they
remember your name. Be friendly!
\end{quote}

\begin{quote}
At this point, the table is filled. Another person arrives. Instead of
that person starting a new table, the TF gives the new student their
spot and moves their belongings to a new table. No student ever sits
alone. The TF hovers around the table until more students arrive and
start filling out table \#2. And so on. At each stage, students are
responsible for, at a minimum, introducing themselves to the TF and,
even better, to the other students. Best is when students who are
already present shower newly arriving students with welcomes and
introductions.
\end{quote}

All students benefit from your efforts to create a community around your
class. But the students who benefit the most are the ones least likely
to have a community of their own. Popular, sociable students will always
have someone to study with, someone to work on the problem sets with.
Shy students, those with fewer friends and worse social skills, love
Study Halls because the structure ensures that there will always be a
place for them. They will be welcomed because we have created a
community in which being welcoming is a requirement.

\hypertarget{final-projects}{%
\subsection*{Final projects}\label{final-projects}}
\addcontentsline{toc}{subsection}{Final projects}

Research projects in statistics and data science classes often work
well.\footnote{See Ledolter (1995), Wardrop (1999), and White (2018) for
  discussion about the use of projects.} Rotating one-on-one
presentations (ROOOP) can work in any class in which students create a
final project. The only necessary requirement is something to show,
something around which to center the discussion.

The mechanics of the process are outlined in this example e-mail to
students, interspersed with my comments.

\begin{quote}
Below are details on the process for Demo Day. But, really, don't sweat
it. Everything just sort of works out. Just make sure you bring your
(fully charged) laptop. Do not arrive late or points will be deducted.
\end{quote}

My framing is intended to minimize student stress, to make the event
fun. Calling it ``Demo Day'' highlights the connection to the
non-academic world, a connection which my courses try to cultivate and
which students appreciate. (Unlike us, almost all of them will leave
academia.) The two most important logistical issues are student laptop
readiness and an on-time start, so I mention both in the opening
paragraph, the only part of the e-mail which I am confident most
students will read.

\begin{quote}
Main purpose of Demo Day is to get feedback from your peers so that you
can use the next 10 days to make your final submission even better.
\end{quote}

Although Demo Day is graded, the final version of student projects are
not due for another week or so. Without ``guidance,'' students will
often not start to work on their projects till the last possible minute.
By having Demo Day so far in advance of the final due date, we
enable/force students to spread out the work on their projects.

Student presentations themselves are not graded. First, doing so is
stressful to students. Second, it is hard for course staff to ``see''
every presentation, at least from start to finish. Third, because there
are so many more students than staff, we inevitably see some students
during their first presentation and other students for their 7th or 8th.
The latter are much smoother and more comfortable than the former,
unsurprisingly.

However, we do grade the quality of the code and the other materials
associated with the presentation. This forces students to have completed
their projects, even though they have another week or more before the
final version is due.

\begin{quote}
Arrive a few minutes early. We start on time! If, for some reason, you
need to present in the first slot, arrive 15 minutes early. Once you
arrive, put your stuff (backpack, coat) off to the side of the room.
Print your name clearly on the sign up sheet at the front of the room.
\end{quote}

In a class with 20 students, mechanics are easy. With scores, even
hundreds, of students, details matter. You need a mechanism for keeping
track of which students actually showed up. You need to plan for
movement around the room.

The bigger the room you can use, the better. The process does work,
however, even if you are in a small room with students all presenting
next to each other, sitting around a seminar table. Just have them keep
their voices down.

\begin{quote}
Students are split into two groups: A and B. The A group starts as
``presenters.'' Grab your computer and sit down one seat in from the
edge of the aisle. Spread out around the room, not too near anyone else.
Bring up your website. Load up your GitHub repo in a browser tab.
\end{quote}

In my introductory data science class, all students complete individual
projects using R and Quarto. The final product is a Quarto website
featuring a few pages with graphics and analysis.

\begin{quote}
Members of group B will select a seat next to a presenter. We rotate. It
doesn't matter where you start. Introduce yourself! Chat with your new
friend.
\end{quote}

If we have an even number, there should be one listener next to each
presenter. If there are an odd number, we will have one extra presenter
in group A. That person will just sit quietly during the first round.

\begin{quote}
A bell anounces the start of the first round. The presenter starts with
their four sentence elevator pitch about their project. Then the
listener asks questions, about anything they want! Maybe they want to
look at the code on GitHub to see how an effect was created. Maybe they
want to talk about the model. Maybe they want a tour of the data
cleaning code. Maybe they just want to poke around the data. Whatever
they like!
\end{quote}

I require students to write and then memorize a four sentence opening
summary about their projects. The world is filled with busy people. If
you want them to spend time with your work, you need to give them a
smooth, coherent case for doing so.

\begin{quote}
A bell, rung after 4 minutes, announces the end of the round. The
presenters stay where they are. The listeners get up, and move on to the
next presenter. (Pay attention to the flow of people around the room so
that you know where to go next.) The bell goes off and we start another
round. This may all seem complex but it just naturally works.
\end{quote}

In practice, we generally don't end up using a bell. Instead, I,
standing at the front of the room and raising my voice, yell ``Time for
all listeners to stand up and move on to the next presentation!'' Even
with that drama, students often need to be shimmied along. On the one
hand, this is nice to see. They are so engaged with the current
presentation that they don't want to leave it. But, like a game of
musical chairs (without the missing chair), student N needs student N+1
to move before she can sit down. Once the next student has arrived, the
presenter begins.

Organizing the movement around the room is more difficult than you might
expect because students don't always pay attention to where they are
supposed to go next.

\begin{quote}
After 8 rounds, the groups switch. B will now present and A will listen.
A puts away their computers and leaves the room to allow B time to set
up. B students go to the front of the room and (anonymously) write down
the names of one or two members of A who created impressive websites
and/or gave nice presentations. (These will not affect our grades, but
we will let students know if their peers thought they did excellent
work.) B students then get their computers and set up, just as A
students did. Students from A then come back in the room and sit down
next to a presenter. Bell goes off and the presentations start again.
\end{quote}

\begin{quote}
After another 8 rounds, we are done. B students leave the room. A
students go to the front and write down their favorite presentations. B
students come back, everyone gathers their stuff, and Demo Day is over.
We finish 75 minutes after we start, just like a regular class session.
\end{quote}

Sadly, we can't assume that students will read this e-mail closely, or
at all.

Rotating one-on-one presentations work virtually, as we found out in
2020. The basic structure is the same on-line as it is in-person, and
for all the same reasons. The more that students speak, the more that
they get out of the session. When listening, the fewer other listeners,
the more that they will pay attention.

One possible modification would be to have a single student present to a
small group of other students. I have found this to be a bad idea for
several reasons.

\begin{itemize}
\item
  Students pay close attention to their peers in a one-on-one
  presentation. They won't even try to look at their phones, except
  during the transitions between sessions. That is, sadly, not true in
  even small groups. Fewer listeners mean a more engaged, albeit
  smaller, audience.
\item
  Group presentations allow for fewer presentations by each student.
  Consider a group of 16. With ROOOP, each student presents 8 times. If,
  instead of groups of 2, we used groups of 4, then each student would
  only present 4 times. The more times that a student is allowed to
  present her work, the better.
\end{itemize}

Another modification is a ``poster style'' presentation in which
students set up to present, either a physical poster or just with their
laptop, and other students wander around the room, listening to various
presentations. This is better than nothing, but far worse in terms of
creating a community because it does not maximize the number of students
spending one-on-one time with each other. A community is built up from
small group interactions, from a meeting of the minds. Although such
meetings can occur during poster style presentations, they are much less
likely.

\hypertarget{conclusion}{%
\subsection*{Conclusion}\label{conclusion}}
\addcontentsline{toc}{subsection}{Conclusion}

Good teaching begins with community. If your students feel that they are
part of a community, they will work harder and learn more. There is no
single trick which creates a community. Instead, there are one hundred
or so tricks, each of which has a small effect on its own. The most
important of these are knowing names, cold calling, classroom seating, a
shallow learning curve, Recitations, Study Halls and rotating-one-on-one
final project presentations. No charisma required.

\hypertarget{references}{%
\subsection*{References}\label{references}}
\addcontentsline{toc}{subsection}{References}

\hypertarget{refs}{}
\begin{CSLReferences}{1}{0}
\leavevmode\vadjust pre{\hypertarget{ref-LedolterJohannes1995PiIS}{}}%
Ledolter, J. (1995),
{``\href{http://www.tandfonline.com/doi/abs/10.1080/00031305.1995.10476184}{Projects
in introductory statistics courses},''} \emph{The American
Statistician}, Taylor \& Francis Group, 49, 364--367.

\leavevmode\vadjust pre{\hypertarget{ref-Lemov2015}{}}%
Lemov, D. (2015), \emph{Teach like a champion 2.0 : 62 techniques that
put students on the path to college}, San Francisco: Jossey-Bass.

\leavevmode\vadjust pre{\hypertarget{ref-gradescope}{}}%
Singh, A., Karayev, S., Gutowski, K., and Abbeel, P. (2017),
{``Gradescope: A fast, flexible, and fair system for scalable assessment
of handwritten work,''} in \emph{Proceedings of the fourth (2017) ACM
conference on learning @ scale}, L@s '17, New York, NY, USA: Association
for Computing Machinery, pp. 81--88.
\url{https://doi.org/10.1145/3051457.3051466}.

\leavevmode\vadjust pre{\hypertarget{ref-wardrop_1999}{}}%
Wardrop, R. L. (1999),
{``\href{http://pages.stat.wisc.edu/~wardrop/papers/tmoore.pdf}{Small
student projects in an introductory statistics course}.''}

\leavevmode\vadjust pre{\hypertarget{ref-prairielearn}{}}%
West, M., Herman, G. L., and Zilles, C. (2015), {``PrairieLearn:
Mastery-based online problem solving with adaptive scoring and
recommendations driven by machine learning,''} in \emph{2015 ASEE annual
conference \& exposition}, pp. 26--1238.

\leavevmode\vadjust pre{\hypertarget{ref-white2018project}{}}%
White, D. (2018), \emph{\href{https://arxiv.org/abs/1802.08858}{A
project based approach to statistics and data science}}.

\end{CSLReferences}

\end{document}